\documentclass[aps,preprint]{revtex4}%
\usepackage{amsfonts}
\usepackage{amsmath}
\usepackage{amssymb}
\usepackage{graphicx}%
\begin{document}
\preprint{ }
\title[Network traffic behaviour near phase transition point]{Network traffic behaviour near phase transition point}
\author{Anna T. Lawniczak}
\affiliation{Department of Mathematics and Statistics \& Guelph-Waterloo Physics Institute,
The Biophysics Interdepartmental Group (BIG), University of Guelph, Guelph,
Ontario N1G 2W1, Canada}
\author{Xiongwen Tang}
\affiliation{Department of Statistics and Actuarial Science, University of Iowa, Iowa City,
Iowa 52242-1409, USA}

\begin{abstract}
We explore packet traffic dynamics in a data network model near phase
transition point from free flow to congestion. The model of data network is an
abstraction of the Network Layer of the OSI (Open Systems Interconnection)
Reference Model of packet switching networks. The Network Layer is responsible
for routing packets across the network from their sources to their
destinations and for control of congestion in data networks. Using the model
we investigate spatio-temporal packets traffic dynamics near the phase
transition point for various network connection topologies, and static and
adaptive routing algorithms. We present selected simulation results and
analyze them.

\end{abstract}
\volumeyear{year}
\volumenumber{number}
\issuenumber{number}
\eid{identifier}
\date[November 01, 2005]{}
\startpage{1}
\endpage{ }
\maketitle

\section{Introduction}

A \emph{Packet Switching Network} (PSN) is a data communication network
consisting of a number of nodes (i.e., routers and hosts) that are
interconnected by communication links. Its purpose is to transmit messages
from their sources to their destinations. In a PSN each message is partitioned
into smaller units of information called \emph{packets} that are capsules
carrying the information payload. Packets are transmitted across a network
from source to destination via different routes. Upon arrival of all packets
to the destination, the message is rebuilt. PSNs owe their name to the fact
that packets are individually switched among routers. Some examples of PSNs
are the Internet, wide area networks (WANs), local area networks (LANs),
wireless communication systems, ad-hoc networks, and sensors networks. There
is vast engineering literature devoted to PSNs, see \cite{Stallings1998},
\cite{Sheledon2001}, \cite{LeonGarcia2000} and the references therein. Wired
PSNs are described by the ISO (International Standard Organization) OSI (Open
Systems Interconnect) 7 layers Reference Model \cite{Sheledon2001},
\cite{LeonGarcia2000}. We focus on the Network Layer because it plays the most
important role in the packet traffic dynamics of the network. It is
responsible for routing packets from their sources to their destinations and
for control of congestion.

The dynamics of flow of packets in PSN can be very complex. It is not well
understood how these dynamics depend on network connection topology coupled
with various routing algorithms. Understanding of packet flow dynamics is
important for further evolution of PSNs, improvements in their design,
operation and defence strategies. Some aspects of these dynamics can be
studied using models of data networks. With different goals in mind and at
various levels of abstractions, models of PSNs have been proposed and applied
(e.g., \cite{Stallings1998}, \cite{Filipiak1988}, \cite{Bertsekas1992},
\cite{Serfozo1999}, \cite{OhiraSawatari1998}, \cite{FuksLawniczak1999},
\cite{LawniczakGerischDiStefano2003}, \cite{LawniczakGerischDiStefano2005},
\cite{YuanMills2003}, \cite{ArrowsmithMondragWoolf2005},
\cite{KocarevVattay2005} and the articles therein).

In our work we explore packet traffic dynamics near phase transition point
from free flow to congestion. We study how this dynamics is affected by the
coupling of network connection topology with routing algorithms. We consider
static and adaptive routing. Continuing our work of
\cite{LawniczakGerischDiStefano2005}, \cite{LawniczakMaxieGerisch2004ACRI},
and \cite{MaxieLawniczakGerisch2004}, we use an abstraction of the Network
Layer (see \cite{LawniczakGerischDiStefano2003},
\cite{LawniczakGerischDiStefano2005}) and a C++ simulator, called Netzwerk-1
(see \cite{LawniczakGerischDiStefano2005}, and
\cite{GerischLawniczakDiStefano2003}) that we developed. Like in real networks
our model is concerned primarily with packets and their routing; it is
scalable, distributed in space, and time discrete. It avoids the overhead of
protocol details present in simulators designed with different goals in mind.

\section{Packet Switching Network Model Description}

\label{PSNModel}

In our PSN model (see, \cite{LawniczakGerischDiStefano2003},
\cite{LawniczakGerischDiStefano2005}) each message consists only of one packet
carrying only the following information: time of creation, destination
address, and number of hops taken. Each node can perform the functions of a
\emph{host} and a \emph{router.} Packets are created randomly and
independently at each node. The probability $\lambda$ with which packets are
created is called \emph{source load}. Each node maintains one incoming and one
outgoing queue to store packets. The outgoing queues are of unlimited length
and operate in a first-in, first-out manner. Each node at each time step
routes the packet from the head of its outgoing queue to the next node on its
route independently from the other nodes. A discrete time, synchronous and
spatially distributed network algorithm implements the creation and routing of
packets \cite{LawniczakGerischDiStefano2003},
\cite{LawniczakGerischDiStefano2005}.

We view a PSN connection topology as a weighted directed multigraph
$\mathcal{L}$ where each node corresponds to a vertex and a pair of parallel
edges oriented in opposite directions represents each communication link. We
associate a packet transmission cost to each directed edge, thus parallel
edges do not necessarily share the same cost. Here we consider network
connection topologies of the type $\mathcal{L}$ = $\mathcal{L}_{\square}%
^{p}(L,l),$ that is, isomorphic to a two-dimensional periodic square lattice
with $L$ nodes in the horizontal and vertical directions and $l$ additional
links added to this square lattice. Notice, that if a sufficient number
$l=l_{1}+l_{2}$ of additional links is added to $\mathcal{L}$ = $\mathcal{L}%
_{\square}^{p}(L,0),$ with $l_{1}$ links added in a proper way, then one can
obtain a network connection topology of the type $\mathcal{L}_{\square}%
^{p}(L,l_{1}+l_{2})=\mathcal{L}_{\triangle}^{p}(L,l_{2}),$ that is isomorphic
to a two-dimensional periodic triangular lattice with $L$ nodes in the
horizontal and vertical directions and $l_{2}$ additional links added to this
triangular lattice. All links in the network are static for the duration of
each simulation run.

In the PSN model, each packet is transmitted via routers from its source to
its destination according to the routing decisions made independently at each
router and based on a least-cost criterion. We consider routing decisions
based on the \emph{minimum least-cost criterion,} that is, \emph{minimum route
distance} or the \emph{minimum route length }depending on the cost assigned to
each edge of the graph \cite{Stallings1998}, \cite{Bertsekas1992}. We consider
the following \emph{edge cost functions} called One ($ONE$), QueueSize ($QS$),
or QueueSizePlusOne ($QSPO$) \cite{LawniczakGerischDiStefano2003},
\cite{LawniczakGerischDiStefano2005}. In each PSN model set-up all edge costs
are computed using the same type of edge cost function.

Edge cost function $ONE$ assigns a value of one to all edges in the lattice
$\mathcal{L}$. This results in the \emph{minimum hop routing }(minimum route
distance) when a least cost routing criterion is applied, because the number
of hops taken by each packet between its source and its destination node is
minimized. This type of routing is called \emph{static routing}, since the
values assigned to each edge do not change during the course of a simulation.
The edge cost function $QS$ assigns to each edge in the lattice $\mathcal{L}$
a value equal to the length of the outgoing queue at the node from which the
edge originates. When this edge cost function is used a packet traversing a
network will travel from its current node to the next node along an edge
belonging to a path with the least total number of packets in transit between
its current location and its destination at this time. The edge cost function
$QSPO$ assigns a summed value of a constant one plus the length of the
outgoing queue at the node from which the edge originates. This cost function
combines the features of the previous two functions. The costs $QS$ and $QSPO$
are derived for each edge from the router's load. Since the routing decisions
made using $QS$ or $QSPO$ edge cost function rely on the current state of the
network simulation they imply \emph{adaptive} or \emph{dynamic routing.} In
these types of routing packets have the ability to avoid congested nodes
during a network simulation.

Our PSN model uses \emph{full-table routing}, that is, each node maintains a
\emph{routing table} of least path cost estimates from itself to every other
node in the network. Because in a PSN model using edge cost function $ONE$ the
costs of edges are static during a simulation, the routing tables are
calculated only at the beginning, they are not updated, and the cost estimates
are the precise least-costs \cite{LawniczakGerischDiStefano2005}. When the
edge cost function $QS$ or $QSPO$ is used, the routing tables are updated at
each time step using a distributed version of Bellman-Ford least-cost
algorithm \cite{Bertsekas1992}. In both cases the path costs stored in the
routing tables are only estimates of the actual least path costs across the
network because only local information is exchanged and updated at each time
step. In our PSN model time is discrete and we observe its state at the
discrete time points $k=0,1,2,\ldots,T$, where $T$ is the final simulation
time. At time $k=0$, the set-up of PSN model is initialised with empty queues
and the routing tables are computed using the centralised Bellman-Ford
least-cost algorithm \cite{Bertsekas1992}.

The discrete time, synchronous and distributed in space PSN model algorithm
consists of the sequence of five operations advancing the simulation time from
$k$ to $k+1$. These operations are: (1)~\emph{Update routing tables,}
(2)~\emph{Create and route packets, }(3)~\emph{Process incoming queue,}
(4)~\emph{Evaluate network state,} (5)~\emph{Update simulation time.} This
algorithm is described in \cite{LawniczakGerischDiStefano2003},
\cite{LawniczakGerischDiStefano2005}.

\section{Packet traffic behaviour near phase transition point}

\label{PacketTraffic}

The phase transition from congestion-free to congested state was observed in
empirical studies of PSNs \cite{TretyakovTakayasuTakayasu1998} and motivated
further research (e.g., \cite{KocarevVattay2005} and the articles therein).
Understanding the dynamics of this transition has practical implications
leading to more efficient designs of PSNs. In our PSN model, for a particular
family of network set-ups, which differ only in the value of the source load
$\lambda$, values $\lambda$ for which packet traffic is congestion-free are
called sub-critical while values for which traffic is congested are called
super-critical. The critical source load $\lambda_{c}$ is the largest
sub-critical source load. The explanation how we estimate its value in
simulations is given in \cite{LawniczakGerischDiStefano2005}.%

\begin{table}[tbp] \centering
\begin{tabular}
[c]{|l|l|l|l|l|}\hline
& $\mathcal{L}_{\square}^{p}(16,0)$ & $\mathcal{L}_{\square}^{p}(16,1)$ &
$\mathcal{L}_{\triangle}^{p}(16,0)$ & $\mathcal{L}_{\triangle}^{p}%
(16,1)$\\\hline
$\mathbf{ONE}$ & 0.115 & 0.020 & 0.140 & 0.030\\
$\mathbf{QS}$ & 0.120 & 0.125 & 0.155 & 0.160\\
$\mathbf{QSPO}$ & 0.120 & 0.125 & 0.155 & 0.160\\\hline
\end{tabular}
\caption{Critical source load values}\label{TableSourceLoads}%
\end{table}%

Here, we explore how spatio-temporal dynamics of packet traffic in our PSN
model is affected by the network connection topology and edge cost function
type for source loads close to the phase transition point of each PSN model
set-up. As a case study we consider network connection topologies of the type
$\mathcal{L}_{\square}^{p}(L,l)$ and $\mathcal{L}_{\triangle}^{p}(L,l),$ where
$L=16$ and $l=0$ or $1.$ We say that $\mathcal{L}_{\square}^{p}(L,0)$ and
$\mathcal{L}_{\triangle}^{p}(L,0)$, that is, the regular network connection
topologies are \emph{undecorated} and we say that they are \emph{decorated} if
an extra link is added to each of them, that is, they are of the type
$\mathcal{L}_{\square}^{p}(L,1)$ and $\mathcal{L}_{\triangle}^{p}(L,1)$. We
use the following convention if we want to specify additionally what type of
an edge cost function our PSN model is using, namely, $\mathcal{L}_{\square
}^{p}(16,l,ecf)$ and $\mathcal{L}_{\triangle}^{p}(16,l,ecf),$where $l=0$ $or$
$1,$ and $ecf=ONE$, or $QS$, or $QSPO$.%

\begin{table}[tbp] \centering
\begin{tabular}
[c]{|l|l|l|}\hline
$\mathcal{L}_{\square}^{p}(16,0,ONE),k=8000$ & $\mathcal{L}_{\square}%
^{p}(16,0,QS),k=8000$ & $\mathcal{L}_{\square}^{p}(16,0,QSPO),k=8000$\\\hline%
{\parbox[b]{1.4209in}{\begin{center}
\includegraphics[
height=1.5731in,
width=1.4209in
]%
{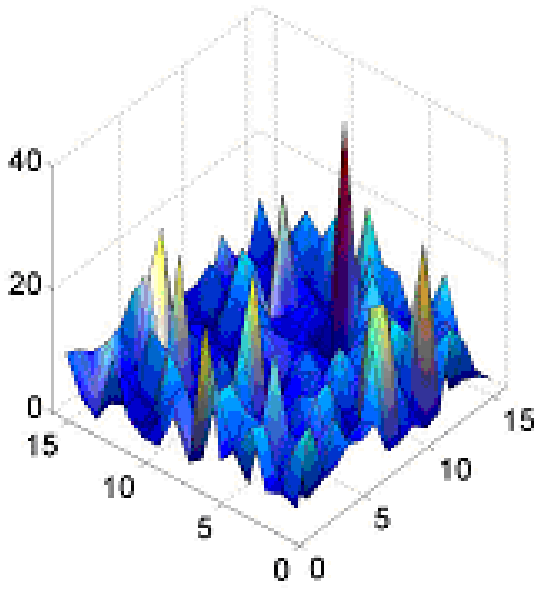}%
\\
$\lambda_{c}=0.115$%
\end{center}}}%
&
{\parbox[b]{1.5134in}{\begin{center}
\includegraphics[
height=1.6155in,
width=1.5134in
]%
{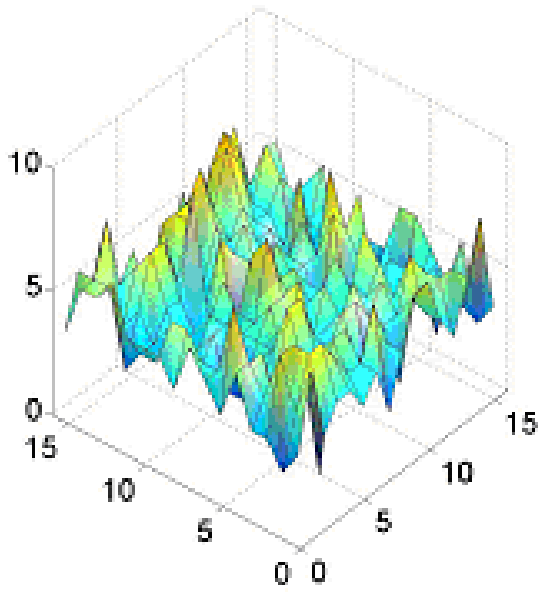}%
\\
$\lambda_{c}=0.120$%
\end{center}}}%
&
{\parbox[b]{1.5203in}{\begin{center}
\includegraphics[
height=1.5273in,
width=1.5203in
]%
{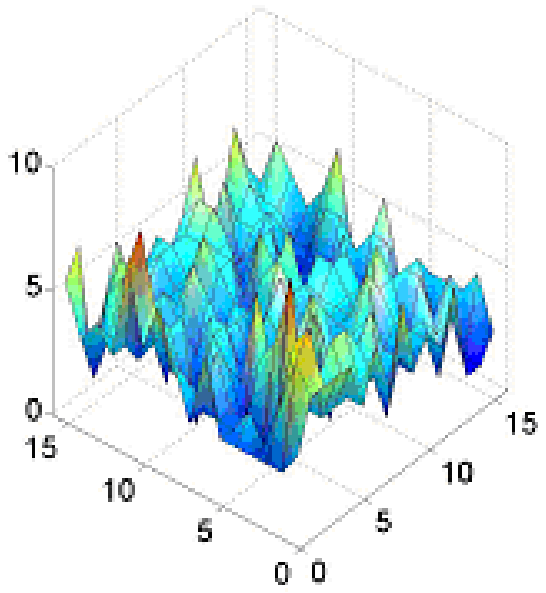}%
\\
$\lambda_{c}=0.120$%
\end{center}}}%
\\\hline%
{\parbox[b]{1.5497in}{\begin{center}
\includegraphics[
height=1.5731in,
width=1.5497in
]%
{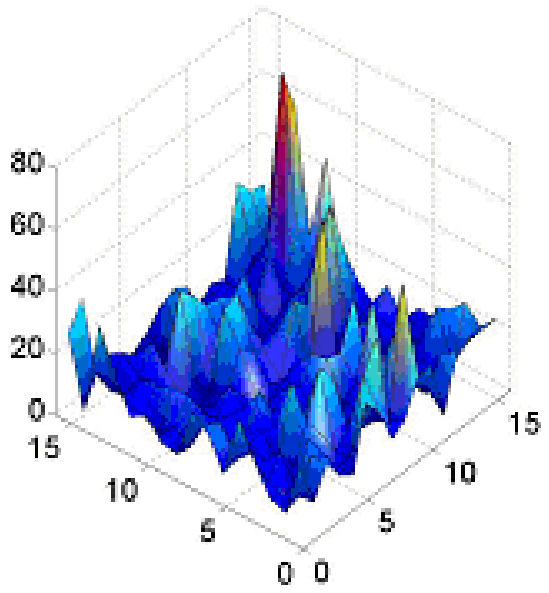}%
\\
$\lambda_{\sup c}=0.120$%
\end{center}}}%
&
{\parbox[b]{1.5618in}{\begin{center}
\includegraphics[
height=1.5264in,
width=1.5618in
]%
{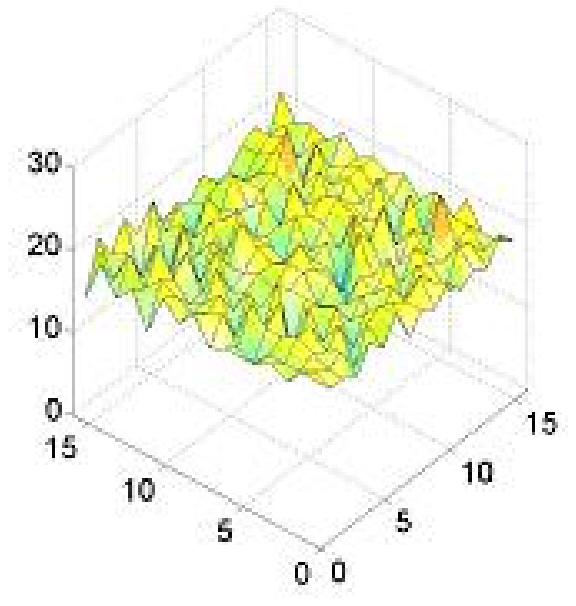}%
\\
$\lambda_{\sup c}=0.125$%
\end{center}}}%
&
{\parbox[b]{1.5618in}{\begin{center}
\includegraphics[
height=1.6328in,
width=1.5618in
]%
{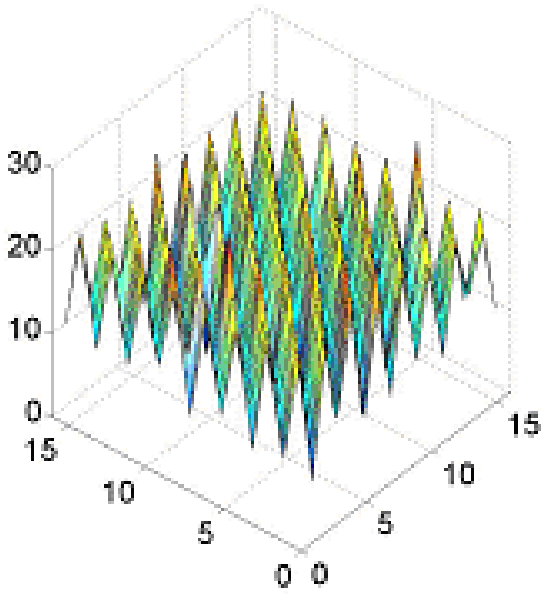}%
\\
$\lambda_{\sup c}=0.125$%
\end{center}}}%
\\\hline
\end{tabular}
\caption{For critical and super-critical source loads spatial distribution of outgoing queue sizes in the PSN model set-ups defined in the table header}\label{TablePeriodicSuareLattice}%
\end{table}%

For the considered network connection topologies the estimated critical source
load values $\lambda_{c}$ are provided in Table \ref{TableSourceLoads}. An
extensive study about how the values of $\lambda_{c}$ depend on the network
connection topology, the edge cost function type, and the mode of routing
table update in the considered PSN model is presented in
\cite{LawniczakGerischDiStefano2005} and reference therein.%

\begin{table}[tpb] \centering
\begin{tabular}
[c]{|c|c|c|}\hline
$\mathcal{L}_{\triangle}^{p}(16,0,ONE),k=8000$ & $\mathcal{L}_{\triangle}%
^{p}(16,0,QS),k=8000$ & $\mathcal{L}_{\triangle}^{p}(16,0,QSPO),k=8000$%
\\\hline%
{\parbox[b]{1.3785in}{\begin{center}
\includegraphics[
height=1.4901in,
width=1.3785in
]%
{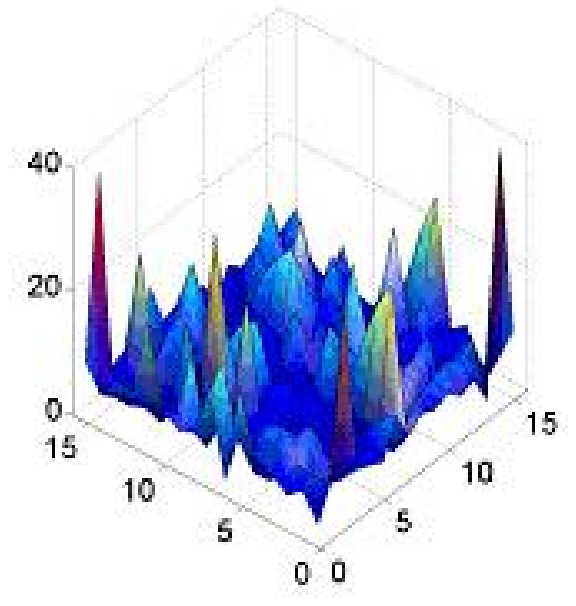}%
\\
$\lambda_{c}=0.140$%
\end{center}}}%
&
{\parbox[b]{1.5497in}{\begin{center}
\includegraphics[
height=1.5437in,
width=1.5497in
]%
{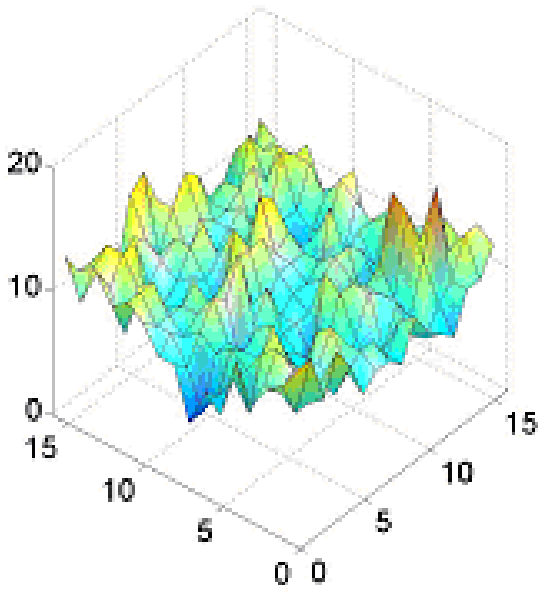}%
\\
$\lambda_{c}=0.155$%
\end{center}}}%
&
{\parbox[b]{1.5428in}{\begin{center}
\includegraphics[
height=1.5324in,
width=1.5428in
]%
{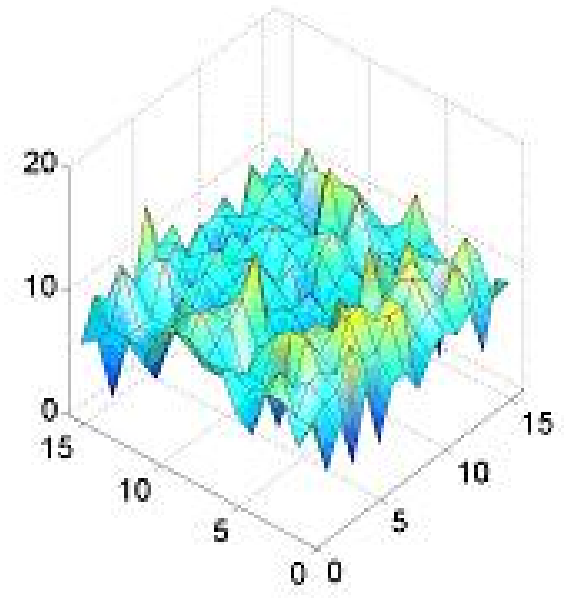}%
\\
$\lambda_{c}=0.155$%
\end{center}}}%
\\\hline%
{\parbox[b]{1.4027in}{\begin{center}
\includegraphics[
height=1.3898in,
width=1.4027in
]%
{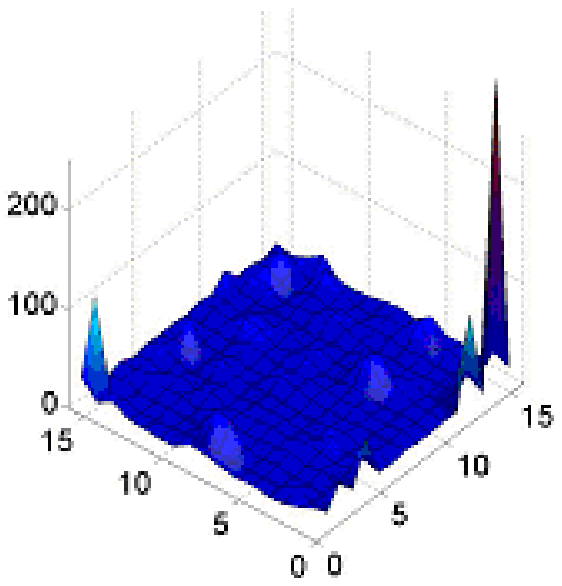}%
\\
$\lambda_{\sup c}=0.145$%
\end{center}}}%
&
{\parbox[b]{1.5618in}{\begin{center}
\includegraphics[
height=1.6397in,
width=1.5618in
]%
{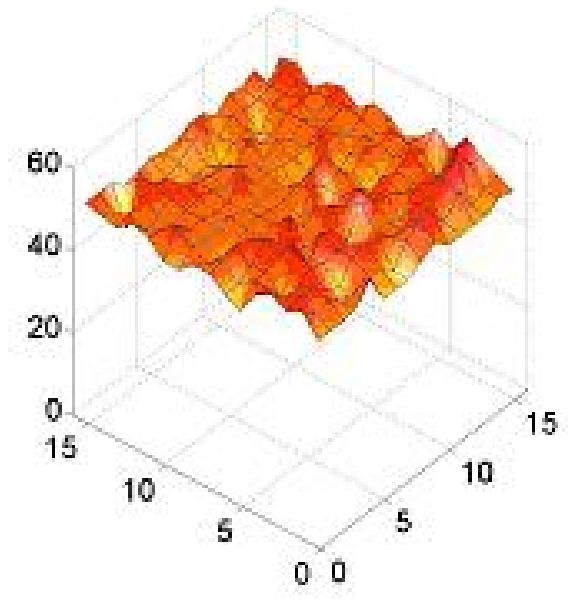}%
\\
$\lambda_{\sup c}=0.160$%
\end{center}}}%
&
{\parbox[b]{1.5428in}{\begin{center}
\includegraphics[
height=1.5506in,
width=1.5428in
]%
{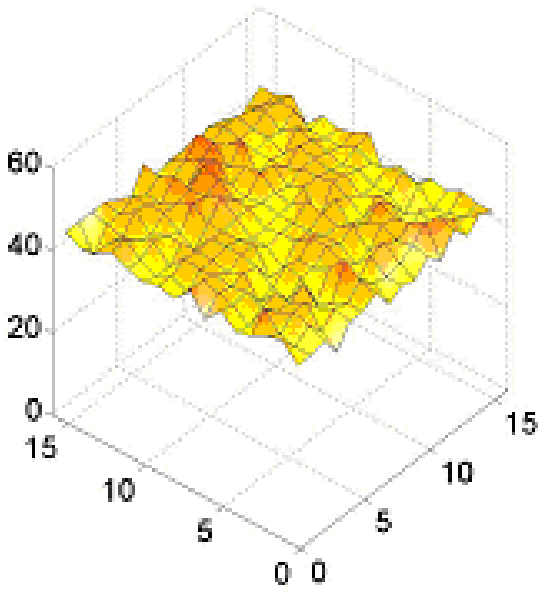}%
\\
$\lambda_{\sup c}=0.160$%
\end{center}}}%
\\\hline
\end{tabular}
\caption{For critical and super-critical source loads spatial distribution of outgoing queue sizes in the PSN model set-ups defined in the table header}\label{TablePeriodicTriangular}%
\end{table}%

Figures of Tables \ref{TablePeriodicSuareLattice},
\ref{TablePeriodicTriangular} and \ref{TableQSDynamics} display spatial
distribution of outgoing queue sizes at nodes for various PSN model set-ups.
The x- and y- axis coordinates of each figure denote the positions of
switching nodes and z-axis denotes the number of packets in the outgoing queue
of the node located at that (x,y) position. The header of each column of each
table and the parameter shown under each figure uniquely define the PSN model
set-up, the simulation time, the values of critical source load $\lambda_{c}$
and super-critical source load $\lambda_{\sup c}=\lambda_{c}+0.005$.

From Table \ref{TablePeriodicSuareLattice} and \ref{TablePeriodicTriangular}
we observe that the qualitative behaviour of spatial distribution of outgoing
queue sizes is very similar for PSN model set-up $\mathcal{L}_{\square}%
^{p}(16,0,ONE)$ and $\mathcal{L}_{\triangle}^{p}(16,0,ONE)$ when the source
loads are $\lambda_{c}$ and $\lambda_{\sup c}$, respectively. In each case
queue sizes are randomly distributed with large fluctuations. Increase in
source load results in substantial increase of queue sizes and fluctuations
among them (notice, the figures use different scales on z-axis). The discussed
behaviours for $\lambda_{c}$ and $\lambda_{\sup c}$ are typically observed for
other sub-critical source load values and super-critical ones, respectively,
in these types of PSN model set-ups.

From Tables \ref{TablePeriodicSuareLattice} and \ref{TablePeriodicTriangular}
we observe also a random distribution of queue sizes in each PSN model set-up
$\mathcal{L}_{\square}^{p}(16,0,ecf)$ and $\mathcal{L}_{\triangle}%
^{p}(16,0,ecf),$ where $ecf=$ $QS$, $QSPO,$ and the source loads are
$\lambda_{c}$. But, the magnitudes of the fluctuations are smaller than those
in the case of PSN model set-ups using the edge cost function $ONE$. For the
other sub-critical source load values in the PSN model set-ups $\mathcal{L}%
_{\square}^{p}(16,0,ecf)$ and $\mathcal{L}_{\triangle}^{p}(16,0,ecf),$ where
$ecf=$ $QS$, $QSPO$, queue size distributions share the same characteristics.
However, we observe a big qualitative difference between the distribution of
queue sizes of the PSN model set-up $\mathcal{L}_{\square}^{p}(16,0,QSPO)$ and
those of the set-ups $\mathcal{L}_{\square}^{p}(16,0,QS),$ $\mathcal{L}%
_{\triangle}^{p}(16,0,QS),$ and $\mathcal{L}_{\triangle}^{p}(16,0,QSPO)$, when
their respective $\lambda_{\sup c}$ source loads are used, see Table
\ref{TablePeriodicSuareLattice} and Table \ref{TablePeriodicTriangular}.

In each of the PSN model set-up $\mathcal{L}_{\square}^{p}(16,0,QSPO)$ and
$\mathcal{L}_{\square}^{p}(16,0,QS)$ for $\lambda_{\sup c}$ (i.e., in the
congested state of the network) we observe spatio-temporal self-organization
in the sizes of the outgoing queues and the emergence of a pattern of peaks
and valleys, see Table \ref{TablePeriodicSuareLattice} and Table
\ref{TableQSDynamics}. However, the time scale on which the pattern emerges in
the PSN model set-up $\mathcal{L}_{\square}^{p}(16,0,QS)$ is much longer than
the time scale of the PSN model set-up $\mathcal{L}_{\square}^{p}(16,0,QSPO)$.
Also, the difference between sizes of the neighbouring peaks and valleys
increases much faster with time in the PSN model set-up $\mathcal{L}_{\square
}^{p}(16,0,QSPO)$ than the one of $\mathcal{L}_{\square}^{p}(16,0,QS)$. This
could imply that, when the edge cost function $QSPO$ is used, the cost
component $ONE$ is responsible for the observed qualitative differences in the
evolution of the spatio-temporal packet traffic dynamics between the network
models under consideration. A similar behaviour was observed for values of $L$
other than $16$ and super-critical source load values other than
$\lambda_{\sup c}$. In the congested state as the number of packets increases
the pattern of peaks and valleys emerges in spite of the adaptive routing
attempts to distribute evenly the packets among the queues.

Looking at Table \ref{TablePeriodicTriangular} and Table \ref{TableQSDynamics}
we see that adding an extra link to the connection topology of the PSN model
set-up $\mathcal{L}_{\square}^{p}(16,0,QS)$ speeds up the \textquotedblleft
peak-valley\textquotedblright\ pattern emergence in a congested state. This is
also true when the extra link has other position and/or length and $L$ is
different than $16$, see \cite{LawniczakGerischDiStefano2005},
\cite{MaxieLawniczakGerisch2004}. The same phenomenon was observed in the
congested states when edge cost function $QSPO$ was used, see
\cite{LawniczakGerischDiStefano2005}, \cite{LawniczakMaxieGerisch2004ACRI}.%

\begin{table}[tpb] \centering
\begin{tabular}
[c]{|l|l|l|}\hline
$\mathcal{L}_{\square}^{p}(16,0,QS),\lambda_{\sup c}=0.125$ & $\mathcal{L}%
_{\square}^{p}(16,1,QS),\lambda_{\sup c}=0.130$ & $\mathcal{L}_{\triangle}%
^{p}(16,1,QS),\lambda_{\sup c}=0.165$\\\hline%
{\parbox[b]{1.5255in}{\begin{center}
\includegraphics[
height=1.6682in,
width=1.5255in
]%
{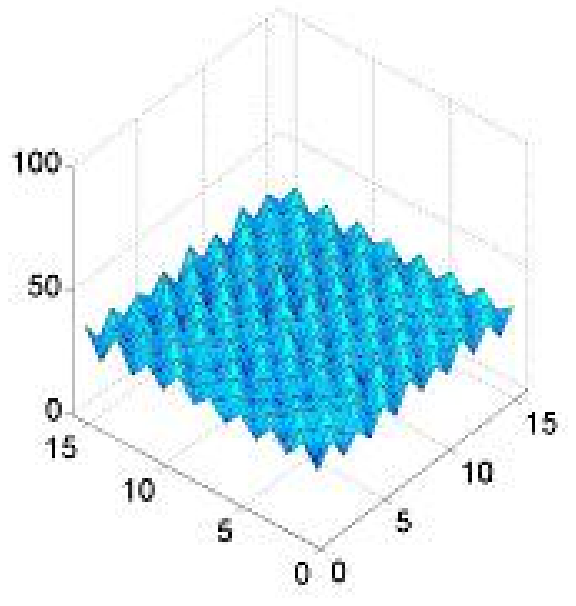}%
\\
$k=20000$%
\end{center}}}%
&
{\parbox[b]{1.4269in}{\begin{center}
\includegraphics[
height=1.5973in,
width=1.4269in
]%
{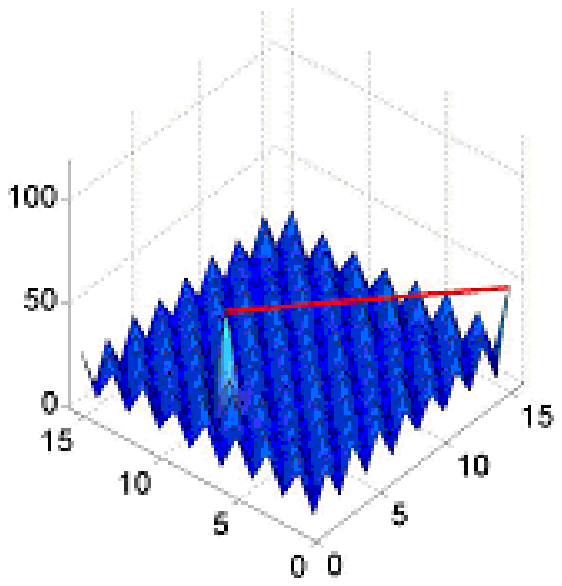}%
\\
$k=2000$%
\end{center}}}%
&
{\parbox[b]{1.5134in}{\begin{center}
\includegraphics[
height=1.5973in,
width=1.5134in
]%
{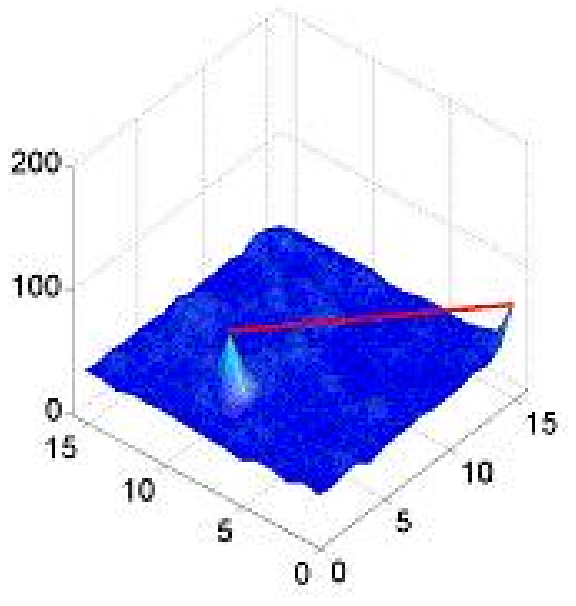}%
\\
$k=2000$%
\end{center}}}%
\\\hline%
{\parbox[b]{1.6051in}{\begin{center}
\includegraphics[
height=1.5324in,
width=1.6051in
]%
{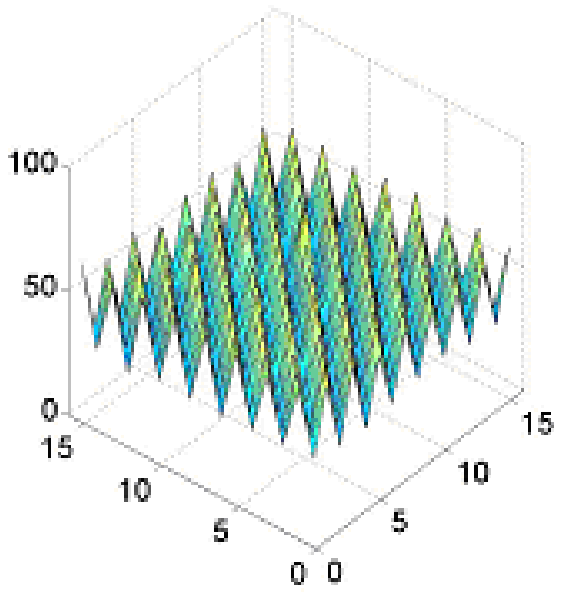}%
\\
$k=40000$%
\end{center}}}%
&
{\parbox[b]{1.4572in}{\begin{center}
\includegraphics[
height=1.5437in,
width=1.4572in
]%
{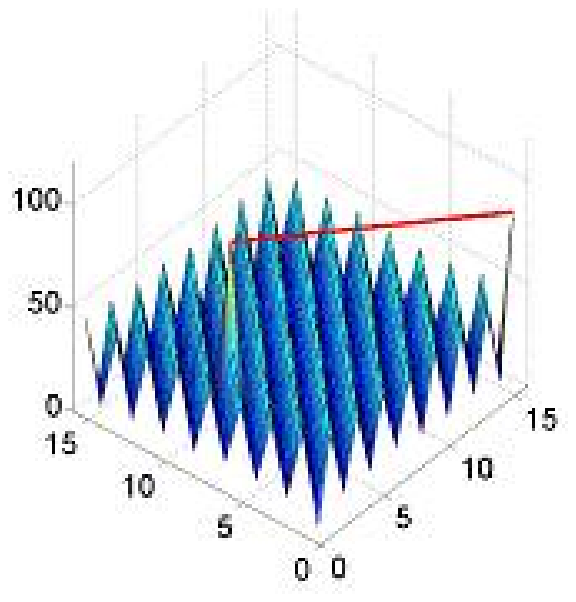}%
\\
$k=4000$%
\end{center}}}%
&
{\parbox[b]{1.5316in}{\begin{center}
\includegraphics[
height=1.6215in,
width=1.5316in
]%
{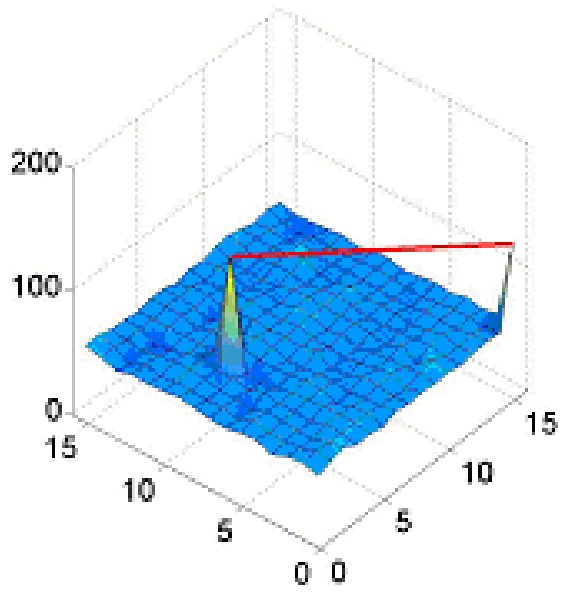}%
\\
$k=4000$%
\end{center}}}%
\\\hline%
{\parbox[b]{1.6103in}{\begin{center}
\includegraphics[
height=1.6397in,
width=1.6103in
]%
{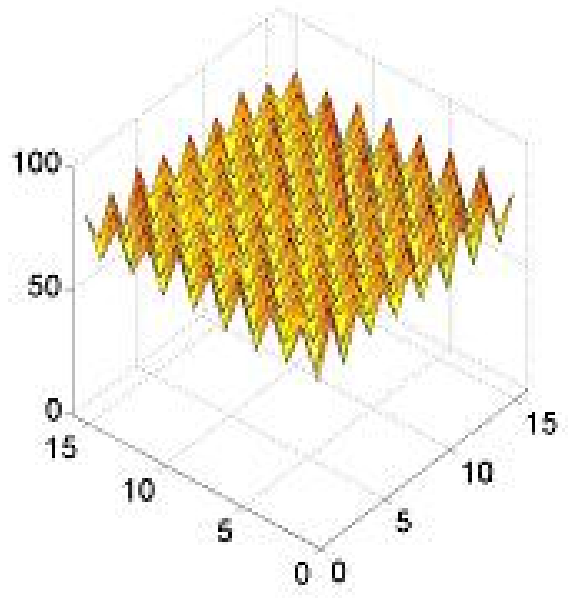}%
\\
$k=80000$%
\end{center}}}%
&
{\parbox[b]{1.5056in}{\begin{center}
\includegraphics[
height=1.484in,
width=1.5056in
]%
{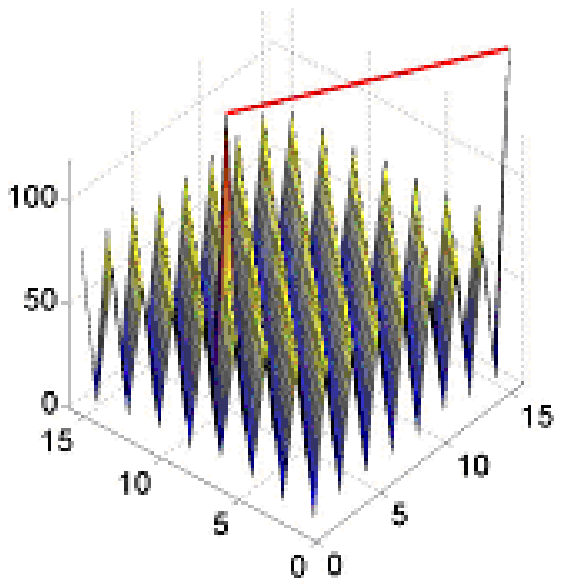}%
\\
$k=8000$%
\end{center}}}%
&
{\parbox[b]{1.58in}{\begin{center}
\includegraphics[
height=1.5627in,
width=1.58in
]%
{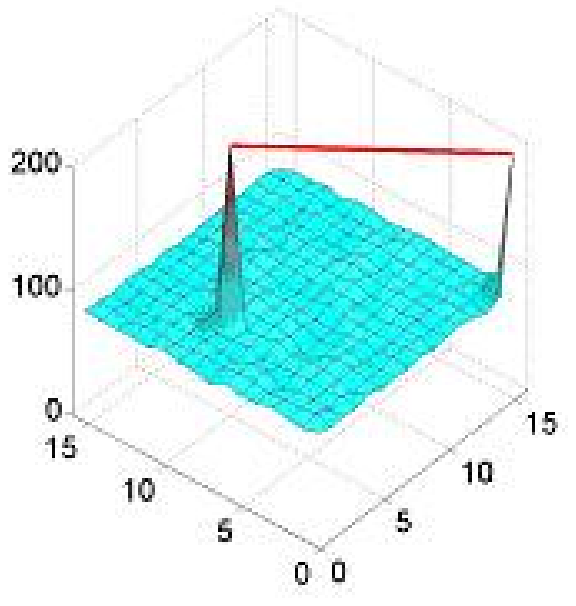}%
\\
$k=8000$%
\end{center}}}%
\\\hline
\end{tabular}
\caption{Spatial distribution of outgoing queue sizes at various times k in the PSN model set-ups defined in the table header }\label{TableQSDynamics}%
\end{table}%

In spite the fact that the considered adaptive routings try to distribute
packets evenly among the network nodes the effect of an extra link on the
packet traffic dynamics is much stronger. It provides a \textquotedblleft
short-cut in communication\textquotedblright\ among distant nodes. This speeds
up the emergence of the \textquotedblleft peak-valley\textquotedblright%
\ pattern in the network models with square lattice connection topologies and
edge cost functions $QS$ and $QSPO$. This is also true when instead of one
extra link a relatively small number of extra links is added, see
\cite{LawniczakGerischDiStefano2005}. If a larger number of extra links is
added the pattern of peaks-valleys does not emerge, see Table
\ref{TablePeriodicTriangular}, Table \ref{TableQSDynamics} and
\cite{LawniczakMaxieGerisch2004ACRI}. We observe rather small differences
among the outgoing queue sizes in congested states of the PSN model set-ups
$\mathcal{L}_{\triangle}^{p}(16,0,QS)$ and $\mathcal{L}_{\triangle}%
^{p}(16,0,QSPO)$. Also, from Table \ref{TableQSDynamics} we notice that when
an extra link is added to the periodic triangular network connection topology,
these differences become even smaller, except of the two nodes to which the
extra link is attached. These nodes attract much larger numbers of packets
than other nodes resulting in local congestion. Qualitatively similar
behaviour was observed for the PSN model set-up $\mathcal{L}_{\triangle}%
^{p}(16,1,QSPO)$ in its congested state.

\section{Conclusions}

\label{Conclusions}

We investigated the effects of coupling of network connection topology with,
respectively, static and adaptive routings on spatio-temporal packet traffic
dynamics near phase transition point from free flow to congested state in our
PSN model. We observed that for adaptive routings and periodic square lattice
network connection topologies patterns of "peaks-valleys" emerge in the
distributions of outgoing queue sizes when the network models are in their
congested states. The emergence of this type of synchronization is accelerated
by addition of an extra link and is destroyed by addition of many links.
Synchronization in other types of networks are discussed in
\cite{MotterZhoyKurths2005}, \cite{Gilbert2005} and references therein. For
our PSN model with adaptive routings and periodic triangular lattice
connection topologies we observed that packet traffic is much more evenly
distributed. We demonstrated that dynamics of the PSN model depend on the
coupling between network connection topology and routing. The presented work
is continuation of our research in \cite{LawniczakGerischDiStefano2005},
\cite{LawniczakMaxieGerisch2004ACRI}, \cite{MaxieLawniczakGerisch2004} and
contributes to the study of dynamics of complex networks.

\begin{acknowledgments}
A.T. L. acknowledges partial financial support from Sharcnet and NSERC of
Canada. X.T. acknowledges partial finantial support from Sharcnet and the
Universty of Guelph. The authors thank to B. Di Stefano, A. Gerisch and K.
Maxie for helpful discussions.
\end{acknowledgments}


\begin{thebibliography}{99}                                                                                               %


\bibitem {Stallings1998}W. Stallings, \emph{High-Speed Networks: TCP/IP and
ATM Design Principles }(Prentice Hall, Upper Saddle River, New Jersey 1998).

\bibitem {Sheledon2001}T. Sheldon, \emph{Encyclopedia of Networking \&
Telecommunications }(Osborne/McGrawHill, Berkeley, California 2001).

\bibitem {LeonGarcia2000}A. Leon-Garcia and I. Widjaja, \emph{Communication
Networks} (McGraw-Hill, Boston, 2000).

\bibitem {Filipiak1988}J. Filipiak, \emph{Modelling and Control of Dynamic
Flows in Communication Networks} (Springer-Verlag Berlin Heidelberg 1988).

\bibitem {Bertsekas1992}P.D. Bertsekas and R.G. Gallager, \emph{Data Networks}
(Prentice Hall, Upper Saddle River 1992).

\bibitem {Serfozo1999}R. Serfozo, \emph{Introduction to Stochastic Networks
}(Springer-Verlag Berlin Heidelberg New York 1999).

\bibitem {OhiraSawatari1998}T. Ohira and R. Sawatari, \emph{Physical Review E}
\textbf{58} (1998) 193-195.

\bibitem {FuksLawniczak1999}H. Fuks and A.T. Lawniczak, \emph{Mathematics and
Computers in Simulations}, \textbf{51}(1999) 101-117.

\bibitem {LawniczakGerischDiStefano2003}A.T. Lawniczak, A. Gerisch and B. Di
Stefano, in \emph{Proceedings in IEEE CCECE} \emph{2003-CCGEI 2003}, Montreal,
Quebec, Canada (2003) 001-004

\bibitem {LawniczakGerischDiStefano2005}A.T. Lawniczak, A. Gerisch and B. Di
Stefano, in \emph{Science of Complex Networks} (J. F. Mendes, Ed., AIP
Conference Proceedings, Vol. \textbf{776}, 2005) 166-200.

\bibitem {YuanMills2003}J. Yuan and K. Mills, \emph{Journal of Research of
NIST}, \textbf{107} (2) (2002) 179-191.

\bibitem {ArrowsmithMondragWoolf2005}D. K. Arrowsmith, R.J. Mondrag and M.
Woolf, in \emph{Complex Dynamics in Communication Networks, }Ed., L. Kocarev
and G. Vattay (Springer-Verlag New York 2005) 127-159.

\bibitem {KocarevVattay2005}L. Kocarev and G. Vattay, \emph{Complex Dynamics
in Communication Networks} (Springer-Verlag New York 2005).

\bibitem {LawniczakMaxieGerisch2004ACRI}A.T. Lawniczak, K. P. Maxie and A.
Gerisch, in \emph{Lecture Notes in Computer Science}, Vol. 3305
(Springer-Verlag 2004) 325-334.

\bibitem {MaxieLawniczakGerisch2004}K.P. Maxie, A.T. Lawniczak and A. Gerisch,
in \emph{Proceedings IEEE CCECE 2004-CCGEI} \emph{2004}, Niagara Falls,
Ontario, Canada (2004) 2425-2428.

\bibitem {GerischLawniczakDiStefano2003}A. Gerisch, A.T. Lawniczak and B. Di
Stefano, in \emph{Proceedings IEEE CCECE} \emph{2003-CCGEI 2003}, Montreal,
Quebec, Canada (2003) 001-004.

\bibitem {TretyakovTakayasuTakayasu1998}A. Y.Tretyakov, H. Takayasu and M.
Takayasu, Physica A \textbf{253}, 315 (1998).

\bibitem {MotterZhoyKurths2005}A.E. Motter, C. Zhou and J. Kurths, in
\emph{Science of Complex Networks} (J. F. Mendes, Ed., AIP Conference
Proceedings, Vol. \textbf{776}, 2005) 201-214.

\bibitem {Gilbert2005}A. Gilbert, in \emph{Complex Dynamics in Communication
Networks, }Ed., L. Kocarev and G. Vattay (Springer-Verlag New York 2005) 21-47.
\end{thebibliography}
\end{document}